\documentclass[12pt,a4paper]{article}
\usepackage[utf8]{inputenc}
\usepackage{times}
\usepackage{float}
\usepackage{geometry}
\usepackage{setspace}
\usepackage{caption}    
\usepackage{titling}
\usepackage{ragged2e}
\usepackage{amssymb}
\usepackage{amsmath} 
\usepackage{hyperref}

\title{\textbf{Virtual Garbage Collector (VGC): A Zone-Based Garbage Collection Architecture for Python's Parallel Runtime}}

\author{
\textbf{Abdulla M} \\
\small Email: \texttt{kmabdullah678@gmail.com} \\
\small Source Code: \url{https://github.com/Abdullahlab-n/VGC-for-arxiv}
}

\date{}

\begin{document}
\maketitle

\begin{abstract}
\noindent
{\raggedright
\textbf{The Virtual Garbage Collector (VGC)} proposes a zone-based memory management architecture aimed at improving execution predictability and memory behavior in Python runtimes. The design explores a dual-layer model consisting of an \emph{Active VGC}, responsible for managing runtime object lifecycles, and a \emph{Passive VGC}, intended as a compile-time optimization layer for static allocation planning.  

Rather than relying on traditional heap traversal or generational heuristics, VGC introduces memory zoning and checkpoint-based state evaluation to reduce allocation churn and constrain garbage collection scope. Execution partitioning is experimentally evaluated to isolate workloads and localize memory pressure, enabling more deterministic behavior under loop-intensive, recursive, and compute-heavy workloads.  

This work presents the architectural principles, execution model, and experimental observations of VGC within a partition-aware runtime context. While the full realization of the dual-layer design is an ongoing effort, the results indicate that zone-based allocation and partitioned execution provide a viable foundation for improving scalability and memory predictability in Python-oriented systems.\par}
\end{abstract}

\section{Introduction}
{\raggedright
Garbage collection is a fundamental component of modern programming languages, automating memory management to prevent resource leaks and improve software reliability. In Python, memory management relies heavily on reference counting and the Global Interpreter Lock (GIL)\cite{PythonGIL}, which together introduce several well-known limitations, including increased memory overhead, fragmentation, non-deterministic collection pauses, and restricted scalability for CPU-bound workloads. These constraints become more pronounced in data-intensive, loop-heavy, and multithreaded execution scenarios.

Classical garbage collection techniques typically rely on heap traversal, reference counting, or generational heuristics, all of which incur runtime overhead and introduce variability in pause behavior \cite{Wilson1992}. While such approaches are effective for general-purpose workloads, they provide limited control over object lifecycles and memory locality, particularly in high-performance and parallel execution contexts.

To explore alternative design choices, this paper proposes the \textbf{Virtual Garbage Collector (VGC)}, a zone-based memory management framework designed to improve execution predictability and memory behavior in Python-oriented runtimes. VGC introduces a dual-layer architectural model consisting of an \textbf{Active VGC}, responsible for managing runtime object lifecycles, and a \textbf{Passive VGC}, intended as a compile-time mechanism for guiding static allocation decisions. The separation of responsibilities allows memory behavior to be reasoned about more explicitly across different execution phases.

At the core of VGC is a memory zoning strategy that classifies objects into three predefined zones—\textbf{Red}, \textbf{Green}, and \textbf{Blue}—based on observed access patterns and expected computational characteristics. Frequently accessed objects are assigned to the Green zone, while less frequently accessed or higher-cost objects are placed in the Red or Blue zones. This zoning model aims to localize memory pressure, improve cache behavior, and constrain the scope of garbage collection operations.

VGC further introduces a \textbf{checkpoint-based bitfield tracking mechanism} that encodes object state information compactly, enabling constant-time state evaluation using basic bitwise logic operations (e.g., AND, OR, and NOT). By replacing recursive heap traversal with checkpoint-based evaluation, the approach seeks to reduce garbage collection overhead and improve determinism. Additionally, a lightweight \textbf{Yield Memory} execution path is introduced to handle ephemeral operations—such as simple variable assignments and arithmetic expressions—without engaging full garbage collection logic.

This paper presents the architectural design of VGC, its zoning and checkpoint mechanisms, and its integration with a partition-aware execution model. Experimental observations are provided to illustrate the behavior of the proposed system under loop-intensive, recursive, and compute-heavy workloads. While the architecture represents an exploratory step rather than a fully realized replacement for Python’s existing memory management, the results suggest that zone-based allocation and partitioned execution provide a promising foundation for improving scalability and memory predictability in high-level language runtimes.
\par}

\section{System Architecture}
{\raggedright
The \textbf{Virtual Garbage Collector (VGC)} is organized as a layered, zone-oriented architecture designed to promote deterministic memory behavior and controlled execution scalability. Rather than modifying Python’s existing memory manager directly, VGC is structured as an auxiliary architectural framework that explores alternative object lifecycle management strategies alongside partition-aware execution.
At a high level, VGC replaces recursive heap traversal with a compact \textbf{checkpoint-based bitfield representation} for object state tracking. Object liveness and eligibility for reclamation are evaluated using constant-time bitwise operations, reducing reliance on per-object reference inspection during collection phases. This architectural choice aims to constrain garbage collection scope and improve predictability, particularly in loop-intensive and partitioned workloads.
The system architecture separates execution partitioning from memory lifecycle control, allowing garbage collection decisions to be localized within predefined zones rather than applied globally across the heap. This separation provides a conceptual foundation for integrating memory zoning, checkpoint evaluation, and partition-aware execution without imposing changes on Python’s core interpreter semantics.

Detailed mechanisms and operational behavior of the zoning model, checkpoint logic, and execution integration are described in subsequent sections.
\par}

\subsection{Overview of Architectural Layers}

The Virtual Garbage Collector (VGC) is organized into four conceptual layers, each addressing a distinct aspect of object lifecycle management and execution control. The layers are designed to remain loosely coupled, allowing memory management decisions to be reasoned about independently from execution scheduling.

\begin{enumerate}
    \item \textbf{Memory Zoning Layer:}  
    Allocated objects are classified into one of three predefined memory zones—\textbf{Red}, \textbf{Green}, or \textbf{Blue}—based on observed access behavior and expected lifecycle characteristics. The zoning mechanism is intended to localize memory pressure and constrain garbage collection activity by grouping objects with similar usage patterns. Each zone maintains independent allocation metadata, reducing interference between workloads with differing memory access profiles.

    \item \textbf{Checkpoint Bitfield Layer:}  
    Instead of maintaining per-object reference counters, VGC employs a centralized checkpoint-based representation in which object states are encoded using compact bitfields. These bitfields represent coarse-grained lifecycle states (e.g., active, inactive, or eligible for reclamation). Object state evaluation is performed using constant-time bitwise logic operations such as \emph{AND}, \emph{OR}, and \emph{Logical NOT}. This approach avoids recursive heap traversal and supports predictable state evaluation.

    \item \textbf{Yield Memory Layer:}  
    The Yield Memory layer provides a lightweight execution path for ephemeral operations, including temporary variable assignments and short-lived intermediate values. Objects handled within this layer may bypass full garbage collection tracking when their lifetime is limited to a single execution scope. Persistent values may optionally be registered with the checkpoint system once execution completes.

    \item \textbf{Control and Execution Layer:}  
    This layer coordinates execution partitioning and the interaction between memory zones, checkpoint evaluation, and yield memory handling. It encapsulates scheduling decisions and partition-aware execution control while remaining independent of low-level memory reclamation logic. The architecture distinguishes between runtime-oriented management (\emph{Active VGC}) and compile-time allocation guidance (\emph{Passive VGC}), with the latter treated as an intended optimization pathway rather than a fully realized component.
\end{enumerate}

\section{Working Principle}

\subsection{R/G/B Zone Model}
{\raggedright
VGC classifies each allocated object into one of three memory zones based on its
observed access frequency $a(o)$ and mutation rate $\mu(o)$, evaluated through the
checkpoint bitfield state.Unlike traditional generational garbage collectors that rely on object promotion across generations \cite{Lieberman1983}, VGC enforces a static-zone lifecycle in which objects expire and are reallocated rather than migrated.\par}

\begin{itemize}
    \item \textbf{Red Zone (R): Rarely-Accessed Objects.}\\
    Contains objects with low access frequency and low mutation activity. These objects
    contribute minimally to the active working set and are reclaimed aggressively to
    avoid unnecessary memory retention.
    
    \item \textbf{Green Zone (G): Frequently-Accessed Objects.}\\
    Represents the program’s hot working set. Objects in this zone exhibit high
    access frequency, low lookup latency, and are prioritized for fast allocation and
    direct access. This zone is optimized for predictable performance.
    
    \item \textbf{Blue Zone (B): Medium-Access Objects.}\\
    Contains objects with intermediate access or mutation rates. These objects remain
    stable across execution epochs but are not part of the primary active working set.
    Blue-zone memory is compacted periodically and validated through sparse checkpoints.
\end{itemize}

{\raggedright Each zone maintains its own bit-address table, allowing constant-time object indexing
without pointer traversal. This eliminates hierarchical heap walking, reduces cache
misses, and enables deterministic memory access across parallel execution units.
Objects do \textbf{not migrate} between zones; instead, expired objects are reclaimed and
reallocated into their new target zone when required. \par}

\subsection{Yield Memory Execution Path}

For short-lived and ephemeral operations, such as simple variable assignments,
\begin{verbatim}
x = 85
\end{verbatim}
{\raggedright
the standard garbage collection workflow may be intentionally avoided. The \textbf{Yield Memory}
execution path provides a lightweight allocation context in which such operations can be
evaluated without engaging full zoning or checkpoint evaluation mechanisms.

Within this execution path, temporary values are handled in a constrained scope and may be
released immediately after execution completes. By limiting interaction with the broader
memory management pipeline, Yield Memory reduces allocation overhead and avoids unnecessary
garbage collection activity for trivial workloads.

When an object produced within the Yield Memory path must persist beyond its immediate
execution scope, the system notifies the checkpoint mechanism, allowing the object to be
registered and classified according to the active zoning policy. This transition enables
ephemeral execution to coexist safely with longer-lived object management while preserving
predictable memory behavior.\par}

\subsection{Three-Bit Checkpoint Architecture}
{\raggedright
The checkpoint layer constitutes the central mechanism for object state tracking within the
Virtual Garbage Collector (VGC). In contrast to conventional reference counting schemes that
maintain an integer counter for each object, VGC represents object state using a compact,
fixed-width bitfield. This representation is intended to encode coarse-grained lifecycle
information while enabling efficient state evaluation.

The checkpoint model uses a small number of bits to represent object status across evaluation
cycles, allowing lifecycle transitions to be expressed explicitly without relying on recursive
reference updates or heap traversal. Object state transitions are evaluated using constant-time
bitwise logic operations, supporting predictable and bounded checkpoint processing.

By decoupling object liveness evaluation from per-object counters, the checkpoint architecture
provides a structured alternative for exploring lifecycle-aware memory management. Detailed
state semantics and transition rules are described in subsequent sections.\par}

\begin{table}[H]
\centering
\small 
\renewcommand{\arraystretch}{1.3}
\setlength{\tabcolsep}{3pt}
\begin{tabular}{|c|c|p{4.8cm}|p{4.2cm}|}
\hline
\textbf{S.No} & \textbf{Bit State} & \textbf{Meaning} & \textbf{Action} \\
\hline
1 & 000 & Idle (not accessed recently) & Wait / Sleep \\
\hline
2 & 001 & Active (currently in use) & Keep Alive \\
\hline
3 & 010 & Candidate for promotion\newline (higher priority zone) & Evaluate \\
\hline
4 & 011 & Candidate for demotion\newline (lower priority zone) & Evaluate \\
\hline
5 & 100 & Persistent (long-lived object) & Keep Stay \\
\hline
6 & 101 & Deferred (lazy checkpoint,\newline waiting confirmation) & Defer Sweep \\
\hline
7 & 110 & Marked (garbage collection\newline scheduled) & Prepare for\newline Deletion \\
\hline
8 & 111 & Expired (no longer needed) & Reclaim Immediately \\
\hline
\end{tabular}
\caption{Three-Bit Checkpoint State Table showing corresponding meanings and actions.}
\label{tab:checkpoint-states}
\end{table}

\subsection{Object Feature Profiling and Runtime Metrics}

To determine the appropriate zone for each allocated object, VGC evaluates a set of
measurable runtime features. These features reflect how the object behaves during execution
and are updated through lightweight counters maintained by the interpreter and core threads.

\textbf{For an object $o$, the following metrics are recorded:}

\begin{itemize}
    \item $\lambda(o)$ — \textbf{Allocation Rate} (objects/second): The instantaneous rate at which
    objects of the same type are allocated during execution. This is reflected in the short-term intensity of object creation.

    \item $\tau(o)$ — \textbf{Lifetime Duration} (seconds): The elapsed time between the object's
    allocation and its last observed reachability checkpoint.

    \item $\mu(o)$ — \textbf{Mutation Rate} (writes/second): The frequency of state updates to the object,
    indicating how actively the data are being modified.

    \item $a(o)$ — \textbf{Access Rate} (reads/second): The number of read operations referencing the
    object over time, reflecting how frequently it is used.

    \item $s(o)$ — \textbf{Object Size} (bytes): The total allocated memory footprint of the object.

    \item $\rho(o)$ — \textbf{Referential Fan-Out}: The number of distinct inbound references points to
    the object, representing its connectivity within the program's object graph.

    {\raggedright
    \item $\chi(o)$ — \textbf{Computational Complexity Weight}: A dimensionless factor representing the expected cost of evaluating, in a check-point system, or releasing the object during garbage collection.\par}
    
\end{itemize}

{\raggedright These metrics are tracked using exponential moving averages to ensure stability over time, allowing VGC to classify and manage objects deterministically without iterative heap scanning
or recursive reference traversal.\par}

\subsection{Zone Allocation Function}

\begin{itemize}
    \item Let $a(o)$ denote the \textit{access frequency} of object $o$ over a checkpoint window.
    \item Let $\mu(o)$ denote the \textit{mutation rate} of $o$ over the same window.
    \item Let $a_R < a_G$ be the \textit{access thresholds}.
    \item Let $\mu_R < \mu_G$ be the \textit{mutation thresholds}.
    \item Let $C_z(o)$ denote the \textit{cost model} of placing $o$ in zone $z \in \{R, G, B\}$.
\end{itemize}

function of Zone Allocation:

\[
Z(o) =
\begin{cases}
R, & \text{if } a(o) < a_R \ \land\ \mu(o) < \mu_R, \\[4pt]
G, & \text{if } a(o) \ge a_G \ \lor\ \mu(o) \ge \mu_G, \\[4pt]
B, & \text{otherwise, with tie-breaker } \arg\min_{z} C_z(o).
\end{cases}
\]

where $C_z(o)$ is the cost model of placing $o$ in zone $z$.

\subsubsection{Boundary Cases}

\begin{itemize}
    \item At $a(o) = a_R$ or $\mu(o) = \mu_R$, the condition for $R$ is not met (strict inequalities), so the object is not assigned to $R$.
    
    \item At $a(o) = a_G$ or $\mu(o) = \mu_G$, the condition for $G$ is met (non-strict inequalities), so the object is assigned to $G$.
    
    \item If $a_R \le a(o) < a_G$ and $\mu_R \le \mu(o) < \mu_G$, the object falls into the “otherwise” case and is assigned to $B$, except that any ambiguity is resolved by the arg\,min tie-breaker on $C_z(o)$:
    \[
        Z(o) = \arg\min_{z \in \{R, G, B\}} C_z(o).
    \]
\end{itemize}

\subsection{Allocation Decision Function}

To determine the appropriate memory zone for an object $o$, VGC evaluates its runtime features:
\[
\lambda(o),\ \tau(o),\ \mu(o),\ a(o),\ s(o),\ \rho(o),\ \chi(o)
\]
 zone-selection thresholds:
\[
\Theta = \{\tau_R, \tau_G,\ \mu_R, \mu_G,\ a_R, a_G,\ s_R, s_G\}
\]
with the ordering constraints:
\[
\tau_R < \tau_G,\quad \mu_R > \mu_G,\quad a_R > a_G,\quad s_R < s_G.
\]

\[
Z(o) =
\begin{cases}
R, & \tau(o) \le \tau_R \land \mu(o) \ge \mu_R \land a(o) \ge a_R \land s(o) \le s_R, \\[4pt]
G, & \tau_R < \tau(o) \le \tau_G \land \mu_G \le \mu(o) < \mu_R \land a_G \le a(o) < a_R \land s_R < s(o) \le s_G, \\[4pt]
B, & \tau(o) > \tau_G \lor \mu(o) < \mu_G \lor a(o) < a_G.
\end{cases}
\]

\noindent If multiple conditions apply, a tie-breaking rule selects the lowest expected cost zone:
\[
Z(o) = \arg\min_{z \in \{R,G,B\}} C_z(o),
\]
where $C_z(o)$ is the expected per-collection cost of storing the object $o$ in zone $z$ .

\subsubsection{Runtime Features and Thresholds}

For an object $o$, define its measured features:
\[
f(o) = \left( \lambda(o),\ \tau(o),\ \mu(o),\ a(o),\ s(o),\ \rho(o),\ \chi(o) \right).
\]

Let the zone-selection thresholds be
\[
\Theta = \{ \tau_R, \tau_G,\ \mu_R, \mu_G,\ a_R, a_G,\ s_R, s_G \}
\]

with ordering constraints:
\[
\tau_R < \tau_G,\qquad
\mu_R > \mu_G,\qquad
a_R > a_G,\qquad
s_R < s_G.
\]

These constraints enforce monotone separation among the zones.

\subsubsection{Eligibility Predicates}

Define the zone-eligibility predicates:

\[
E_R(o):\ \tau(o) \le \tau_R \ \wedge\ \mu(o) \ge \mu_R \ \wedge\ a(o) \ge a_R \ \wedge\ s(o) \le s_R,
\]

\[
E_G(o):\ \tau_R < \tau(o) \le \tau_G \ \wedge\ \mu_G \le \mu(o) < \mu_R \ \wedge\ a_G \le a(o) < a_R \ \wedge\ s_R < s(o) \le s_G,
\]

\[
E_B(o):\ \tau(o) > \tau_G \ \vee\ \mu(o) < \mu_G \ \vee\ a(o) < a_G \ \vee\ s(o) > s_G.
\]

Define the set of applicable zones:
\[
\mathcal{A}(o) = \{\, z \in \{R, G, B\} \mid E_z(o) \text{ is true} \,\}.
\]

\subsection{Zone Cost Model}

Let $C_{\text{mark}}(o)$ be the cost to mark object $o$, 
$C_{\text{scan}}(o)$ the cost to scan its references, 
and $C_{\text{stage}}(o)$ the cost associated with promotion or demotion 
(i.e., expiration and reallocation rather than in-place relocation).

The per-zone cost is parameterized as:

\[
C_R(o) = 
\underbrace{\alpha_R \chi(o)}_{\text{mark}} +
\underbrace{\beta_R \rho(o)}_{\text{scan}} +
\underbrace{\gamma_R s(o)}_{\text{stage (reclaim)}}
\]

\[
C_G(o) = 
\alpha_G \chi(o) +
\beta_G \rho(o) +
\gamma_G s(o)
\]

\[
C_B(o) = 
\alpha_B \chi(o) +
\beta_B \rho(o) +
\gamma_B s(o)
\]

with policy weights:

\[
\gamma_R > \gamma_G > \gamma_B, \qquad
\alpha_R \approx \alpha_G > \alpha_B, \qquad
\beta_R \ge \beta_G \ge \beta_B,
\]

reflecting that:

\begin{itemize}
\item Red-zone objects are short-lived and frequently reclaimed.
\item Green-zone objects form the stable working set with moderate updates.
\item Blue-zone objects are long-lived and rarely re-staged except during periodic compaction.
\end{itemize}

The expected pause contribution per collection for zone $z$ is:

\[
P_z = \sum_{o \in z} C_z(o) \cdot \pi_z,
\]

{\raggedright where $\pi_z \in (0,1)$ controls the fraction of $C_z(o)$
executed during stop-the-world versus concurrent phases. \par}

\subsubsection{Definitions and Scope}

\begin{itemize}
    \item Let $o$ denote an object.
    
    \item Let $C_{\text{mark}}(o)$ be the cost to mark $o$, 
    $C_{\text{scan}}(o)$ the cost to scan $o$'s references, 
    and $C_{\text{stage}}(o)$ the cost associated with promotion/demotion 
    (i.e., expiration and reallocation rather than in-place relocation).
    
    \item The model per-zone costs using nonnegative parameters and 
    additive feature functions. Units are arbitrary ``work units'' per 
    collection; parameters absorb constant factors of the implementation.
\end{itemize}

\[
C_z(o) = \alpha_z \chi(o) + \beta_z \rho(o) + \gamma_z s(o) \quad \text{for } z \in \{R, G, B\}
\]

\subsubsection{Feature functions}

\begin{itemize}
    \item $\chi(o)$: mark-related work attributable to $o$ (e.g., number of header touches).
    \item $\rho(o)$: scan-related work for $o$'s references (e.g., references traversed).
    {\raggedright
    \item $s(o)$: staging work for $o$ (e.g., bytes or objects expired/reclaimed during (promotion/demotion).\par}
    \end{itemize}
All three are nonnegative and dimensioned in primitive ``units'' that the parameters convert into cost.

\subsubsection{Per-Zone Cost Parameterization}

For a zone $z \in \{R, G, B\}$, let the zone-specific parameters 
$\alpha_{z}, \beta_{z}, \gamma_{z} \ge 0$ control the weighting of the mark, scan, 
and stage components:

\[
C_{z}(o) = \alpha_{z}\,\chi(o) + \beta_{z}\,\rho(o) + \gamma_{z}\,s(o).
\]

\begin{itemize}
    \item The first term $\alpha_{z}\chi(o)$ captures \textbf{mark} complexity.
    \item The second term $\beta_{z}\rho(o)$ captures \textbf{scan} overhead.
    \item The third term $\gamma_{z}s(o)$ captures \textbf{stage cost}, corresponding
    to expiration and reallocation rather than in-place movement.
\end{itemize}

\subsubsection{Estimation Notes}

\begin{itemize}
    \item The parameters $\{\alpha_{z}, \beta_{z}, \gamma_{z}\}$ are typically learned 
    through nonnegative least-squares fitting data from the per-collection profiling data.
    \item Identifiability improves when features $[\chi(o), \rho(o), s(o)]$ vary between
    objects and checkpoint intervals.
    \item To maintain interpretability, feature magnitudes should be scaled to similar 
    ranges, or regularization may be applied.
\end{itemize}

\[
C_{z}(o) = \alpha_{z}\chi(o) + \beta_{z}\rho(o) + \gamma_{z}s(o), \qquad z \in \{R,G,B\}.
\]

\subsection{16-Byte Alignment and VGC Checkpoint Indexing}

{\raggedright CPython aligns all heap objects to a 16-byte boundary due to its internal object layout.Each object begins with a fixed 16-byte header containing the reference count and a pointer
to its type descriptor. Small-object allocations are rounded to the nearest 8 or 16 bytes
by the \texttt{pymalloc} allocator, ensuring consistent alignment across the heap.\par}

{\raggedright This alignment property is critical to VGC. Because objects occupy uniformly aligned memory regions, the Virtual Garbage Collector represents each object using a compact 3-bit entry in
a global checkpoint bitfield. The index of an object in the checkpoint table can therefore
be computed in constant time using integer bit shifts rather than pointer dereferencing or
graph traversal.\par}

{\raggedright
This enables the VGC to perform object state evaluation using logic-gate operations without
scanning the heap, providing predictable and deterministic garbage collection behavior.\par}

\subsubsection{ Zone-Index Mapping and Memory Layout Specification}

To operationalize the checkpoint-based evaluation model, the Virtual Garbage Collector (VGC)
defines a deterministic mapping from physical object addresses to checkpoint bit indices and,
subsequently, to the R/G/B memory zones. This mapping ensures constant-time zone lookup,
predictable garbage-collection behavior, and correct parallel partition assignment.Given that all CPython-managed objects are aligned to 16-byte boundaries, each object's index
in the global checkpoint bitfield is computed as:

\begin{equation}
    \text{Index}(o) = \frac{\text{Address}(o) - \text{Base}}{16}.
\end{equation}

{\raggedright
This index uniquely identifies the object and determines its placement within the zone-specific
bitfield ranges defined below.\par}

\subsubsection{Global Zone Boundary Layout}
\hspace*{8.8em}\text{Let:}
\begin{align*}
  N_R &\ = \text{maximum entries in Red zone}, \\
  N_G &\ = \text{maximum entries in Green zone}, \\
  N_B &\ = \text{maximum entries in Blue zone}.
\end{align*}

VGC partitions the global checkpoint table into three contiguous regions:
\[
\text{CheckpointTable} = [R_{\text{bits}} \mid G_{\text{bits}} \mid B_{\text{bits}}].
\]

The index ranges for each zone are:
\begin{align}
    \text{Red zone}: &\qquad i \in [0,\, N_R), \\
    \text{Green zone}: &\qquad i \in [N_R,\, N_R + N_G), \\
    \text{Blue zone}: &\qquad i \in [N_R + N_G,\, N_R + N_G + N_B).
\end{align}

{\raggedright Thus, zone membership is determined in $O(1)$ time via boundary comparison. The zone
decision formulas in Sections 3.5 and 3.6 determine \emph{which} zone an object belongs to,
while this layout determines \emph{where} the object is placed. \par}

\subsubsection{VGC Adopts a static-zone memory model} 
objects do not migrate between R/G/B zones after
allocation. If checkpoint evaluation indicates that an object should belong to a different zone,
the object undergoes:
\begin{enumerate}
    \item expiration in its current zone.
    \item reallocation into the new zone's index range.
\end{enumerate}

This ensures zone-boundary stability and avoids the complexity of in-place relocation.

\subsubsection{Generational Subdivision Within Zones}

Each zone is subdivided into three generational segments: Gen0, Gen1, and Gen2. For a zone
$z$ of size $N_z$, the index ranges are:

\begin{align}
    \text{Gen0}: &\qquad i \in [\text{start}_z,\; \text{start}_z + \alpha N_z),\\
    \text{Gen1}: &\qquad i \in [\text{start}_z + \alpha N_z,\; \text{start}_z + \beta N_z),\\
    \text{Gen2}: &\qquad i \in [\text{start}_z + \beta N_z,\; \text{start}_z + N_z),
\end{align}

{\raggedright
where typical configuration values are $\alpha = 0.25$ and $\beta = 0.75$. These proportions
are tunable based on workload characteristics.\par}

\subsubsection{Partition Mapping for Parallel Execution}

For parallel execution, each zone $z$ is further divided into $P_z$ partitions. Partition $p$ is
assigned a contiguous index interval:

\begin{equation}
    \text{Partition } p \text{ of zone } z:\quad
    i \in 
    \left[
        \text{start}_z + \frac{p}{P_z} N_z,\;
        \text{start}_z + \frac{(p+1)}{P_z} N_z
    \right).
\end{equation}

{\raggedright This design ensures lock-free operation, isolated memory access per core, and deterministic checkpoint evaluation. The 3-bit entries per object allow boundaries to align with machine-word sizes for SIMD-optimized bitfield processing.\par}

\subsubsection{Integration with Logic-Gate Checkpoint Evaluation}

For any object $o$:

\begin{enumerate}
    \item The address determines its index via Eq.~(1).
    \item The index determines its zone by boundary comparison.
    \item The zone determines:
        \begin{itemize}
            \item the bitfield region storing its 3-bit state,
            \item the corresponding logic-gate evaluation chain (R, G, or B),
            \item the partition---and thus the CPU core---responsible for its processing.
        \end{itemize}
\end{enumerate}

{\raggedright This establishes a complete mapping from object address to zone, generation, partition, and checkpoint evaluation path, all in constant time. \par}

\subsection{Active and Passive VGC Integration}

VGC supports \textbf{Active–Passive dual layering}:
\begin{itemize}
    \item \textbf{Active VGC} — Handles dynamic runtime workloads such as recursion, loops, and asynchronous I/O.
    \item \textbf{Passive VGC} — Handles compile-time or static allocations, aligning objects to stable memory regions.
\end{itemize}

{\raggedright
This separation allows stable management of static objects while enabling rapid adaptation for runtime workloads. The checkpoint layer can differentiate between static and dynamic lifecycles through specific 3-bit states.\par}

\subsection{Parallelism and Partitioning}

VGC integrates a partitioning model that distributes workloads across CPU cores, with zones operating independently. Checkpoint bitfield updates are parallelized, avoiding locking overhead typical in Python’s GIL. Because the checkpoint system is based on 3-bit state logic, partitioning can use \textbf{bitmask operations} to batch-evaluate objects across zones concurrently. This allows multiple threads or interpreters to process memory in parallel without race conditions or serialized refcount updates.
In summary, VGC’s working \textbf{three-bit checkpoint state encoding}, \textbf{zone-based allocation}, \textbf{logic-gate-driven decision making}, \textbf{lambda-optimized yield execution}, and \textbf{parallel execution by partitioning workloads}. This results in a deterministic, low-latency, and highly scalable alternative to Python’s GIL-based garbage collector.

\section{Lambda and Three-Bit Checkpoint Mapping}

Each lambda-evaluated object can be associated with a 3-bit state based on its persistence:
\begin{itemize}
    \item 000 -- ephemeral lambda (immediate discard after evaluation).
    \item 001 -- active lambda (used in a running scope).
    \item 100 -- persistent lambda (checkpointed and stored).
    \item 101 -- deferred evaluation (lazy checkpoint, stored later).
\end{itemize}

\text{Let:}
\begin{itemize}
  \item[] $L$ denote the set of lambda operations,
  \item[] $S$ the set of checkpoint states,
  \item[] $Z$ the set of zones.
\end{itemize} 

\hspace{11.5em}\text{Then:}

\[
\phi : L \to S \to Z
\]

The mapping function \(\phi\) is defined as:
\[
\phi(\lambda x, f(x)) = 
\begin{cases}
    \text{short-lived, no zone allocation} & \text{if } 000 \\
    \text{persistent, mapped to Green zone} & \text{if } 100 \\
    \text{deferred, mapped to Red or Blue zone} & \text{if } 101
\end{cases}
\]

{\raggedright
This mapping ensures that ephemeral tasks do not incur additional checkpoint or zone allocation costs, while persistent and deferred tasks are efficiently integrated into the VGC memory hierarchy
\par}

The use of lambda calculus in VGC therefore provides:
\begin{itemize}
    \item Minimal overhead for small, short-lived computations.
    \item Deterministic behavior through explicit state mapping.
    \item Seamless integration with the 3-bit checkpoint system.
\end{itemize}

\section{Logic Gate–Based Checkpoint Architecture}

The \textbf{Logic Gate–Based Checkpoint Architecture}\cite{Wilson1992}, forms the computational core of the VGC checkpoint system. Rather than relying on iterative reference tracking or recursive object scanning, VGC encodes each object's state within a compact \textbf{3-bit field}, which is processed directly through gate-level logic operations. This design enables constant-time evaluation of object lifecycles and eliminates the overhead associated with conditional branching or traversal-based garbage collection.

\subsection{Checkpoint Logic Integration}

Each allocated object is associated with a 3-bit checkpoint code, represented as:
\[
S = (s_2, s_1, s_0)
\]
{\raggedright
Each bit encodes a specific state flag used to determine object liveness and readiness for reclamation. Rather than relying on traditional numerical reference counting, the proposed system groups checkpoint bitfields into logical blocks and evaluates them using bitwise logic operations. This design enables efficient, low-level state inspection across multiple objects simultaneously.

Object states are interpreted through combinations of logic gates that operate directly on checkpoint bits, allowing rapid binary-level evaluation of allocation status, zone permissions, and lifecycle transitions. The following logic gates are used within the checkpoint evaluation framework:\par}

\begin{itemize}
    \item \textbf{AND gate} (\(A \land B\)) — Ensures that an object is considered live only when both its current state and corresponding zone permission bits are active. This gate enforces strict keep-alive conditions.
    
    \item \textbf{OR gate} (\(A \lor B\)) — Preserves objects that remain associated with higher-priority zones (e.g., Green or Blue) even if their primary activation condition becomes inactive.
    
    \item \textbf{NOT gate} (\(\lnot A\)) — Inverts inactive state bits to identify idle or expired objects that are eligible for reclamation.
    
    \item \textbf{XOR gate} (\(A \oplus B\)) — Detects state transitions by identifying differences between successive evaluation epochs for the same object.
    
    \item \textbf{XNOR gate} (\(\overline{A \oplus B}\)) — Identifies stability across multiple evaluation cycles, preventing premature reclamation of persistent objects.
    
    \item \textbf{NAND} and \textbf{NOR} gates — Used in composite checkpoint circuits to support conditional sweeping and deferred reclamation. These gates are applied to objects marked with specific checkpoint patterns, such as \(101\) (Deferred) and \(110\) (Marked).
\end{itemize}

\subsection{Gate-Level Evaluation Model}

During every checkpoint cycle, the bitfield array is loaded into a low-level logic evaluation routine where each 3-bit pattern undergoes concurrent gate computation. The logic evaluation is expressed as:

\[
O = (S \land Z) \lor (\lnot S \land P)
\]
{\raggedright
where:
\par}
\begin{itemize}
    \item \(S\) represents the current state bits of the object.
    \item \(Z\) represents the zone activation mask (\(R, G, B\)).
    \item \(P\) represents pending or deferred state bits.
    \item \(O\) is the output that determines whether the object is still allocated or released.
\end{itemize}

{\raggedright
This logical combination allows VGC to independently assess the readiness of each object without interacting with neighboring references or recursive memory links. Thus, multiple checkpoint evaluations can occur in parallel with no dependency chain, achieving \(\mathcal{O}(1)\) time complexity per object.\par}

\subsection{Bitfield Processing and Memory Synchronization}

All checkpoint bits are aligned to 16-byte memory boundaries due to the CPython design constraints to ensure cache-line efficiency and predictable access latency. The bitfield is divided into three synchronized regions corresponding to the R, G, and B zones:

\[
\text{Checkpoint Table} = [R_{\text{bits}} \,|\, G_{\text{bits}} \,|\, B_{\text{bits}}]
\]

{\raggedright
Each region is independently evaluated through its own logic chain. For example:\par}
\begin{align*}
R' &= (R \land \lnot B) \lor (R \land G) \\
G' &= (G \lor R) \land \lnot B \\
B' &= (B \land \lnot G)
\end{align*}

{\raggedright These relations prevent inter-zone conflicts and ensure stable memory boundaries across concurrent threads. The bitfield synchronization unit within VGC merges results through gate aggregation, maintaining coherence between Active and Passive VGC layers.\par}

\subsection{Advantages of Gate-Level Checkpoint Processing}

\begin{itemize}
    \item \textbf{Constant-Time Evaluation:} Each logic computation occurs in a fixed number of operations, irrespective of heap size.
    \item \textbf{Parallel-Friendly:} Bitfield arrays can be batch-evaluated across CPU cores without shared reference conflicts.
    \item \textbf{Deterministic Behavior:} Gate-level execution produces consistent results under concurrent workloads.
    \item \textbf{Low Memory Overhead:} Only 3 bits per object are required to encode all active states.
    \item \textbf{Hardware Alignment:} Operations are naturally compatible with CPU bit arithmetic and SIMD instructions.
\end{itemize}

{\raggedright
Through its logic gate–driven checkpoint evaluation, VGC transforms memory management into a design that bridges software-level garbage collection with hardware-inspired parallelism, creating a deterministic and high-performance foundation for Python’s next-generation runtime.\par}

\section{Partition Theory and Parallel Execution Model}

The \textbf{Partition Theory and Parallel Execution Model} of the Virtual Garbage Collector (VGC) establishes a mathematical and architectural foundation for distributing workloads across multiple CPU cores and threads instances. This mechanism enables VGC to achieve deterministic, scalable parallelism while maintaining memory consistency across all active threads and cores. While scalable allocators such as Hoard reduce contention in multithreaded environments \cite{Berger2000}, they operate below the runtime layer and do not address object lifecycle semantics or interpreter-level execution parallelism.

\subsection{Partition and Parallel Execution (PPE) Runtime Architecture}

Partition and Parallel Execution (PPE) is the runtime execution architecture\cite{Tanenbaum2007}, responsible for decomposing workloads into independent partitions and executing them concurrently across available CPU cores and hardware threads. PPE operates independently of memory reclamation logic and serves as the execution backbone upon which the Virtual Garbage Collector (VGC) is layered.

{\raggedright
The primary objective of PPE is to eliminate underutilization of processor resources by ensuring that no CPU core remains idle when parallelizable work is available. Unlike Python’s Global Interpreter Lock (GIL), which serializes bytecode execution, PPE enables true multi-core execution by design.\par}
{\raggedright
PPE targets computational patterns that are traditionally constrained by interpreter-level serialization, including:\par}
\begin{itemize}
    \item Loop-intensive workloads
    \item Recursive and deep-recursive execution
    \item Matrix and numerical computation
    \item Interpreter-driven task dispatch
\end{itemize}

{\raggedright
Each workload is partitioned into logically independent execution units based on the processor’s core and thread topology. These partitions are scheduled for parallel execution using native OS threads with explicit core affinity, ensuring predictable performance scaling across dual-core, quad-core, and higher-core-count systems.\par}
{\raggedright
Formally, PPE maps the total workload \( W \) into \( n \) partitions such that:\par}
\[
W = \bigcup_{i=1}^{n} W_i, \quad W_i \cap W_j = \emptyset \text{ for } i \neq j
\]
where each partition \( W_i \) is executed concurrently on a dedicated core or hardware thread.

{\raggedright
PPE is deliberately lightweight and avoids global synchronization barriers. Inter-partition coordination is restricted to minimal aggregation phases, allowing most execution to proceed without contention. This design enables near-linear scalability as the number of available cores increases.\par}

{\raggedright
Importantly, PPE is orthogonal to garbage collection. While PPE governs \emph{how} computation is executed in parallel, VGC governs \emph{how} memory objects are tracked, reused, and expired. This separation of concerns allows execution scalability and memory efficiency to evolve independently.\par}

\subsection{Core Partitioning Principle}

Partition Theory in VGC defines a mapping between the available CPU cores (\( C \)) and the active Threads (\( T \)) according to the formula:
\[
P = \frac{T}{C}
\]
where \( P \) denotes the partition ratio representing the load distribution per core.  
If \( T > C \), VGC dynamically clusters multiple Threads under a single core using an event-driven scheduler; if \( T \leq C \), interpreters are mapped in a one-to-one configuration to maximize concurrency.  

{\raggedright This partitioning ensures that each thread operates within its isolated memory region, sharing only checkpoint metadata and synchronization bits through the Active–Passive VGC interface.\par}

\subsection{Zone-Based Partition Allocation}

Each R, G, and B zone is subdivided into smaller logical partitions:
\[
Z = \{ R_p, G_p, B_p \}
\]
where \( p \) represents a partition index.  
Each partition executes independently under a dedicated thread or lightweight core instance. The checkpoint and yield layers synchronize periodically to maintain zone integrity without introducing global locking mechanisms.

\begin{itemize}
    \item \textbf{Red Partitions (\( R_p \))} handle complex or long-lived allocations that require deferred reclamation and background logic evaluation.
    \item \textbf{Green Partitions (\( G_p \))} manage frequently accessed or active workloads, providing the lowest latency access paths.
    \item \textbf{Blue Partitions (\( B_p \))} manage transitional or temporary objects that require short-term tracking between creation and collection.
\end{itemize}

{\raggedright The parallel zone partitioning enables fine-grained workload balancing where each partition can execute garbage collection independently without interfering with others.\par}

\subsection{ Parallel  Execution Interpreter Synchronization}

VGC utilizes a lightweight synchronization mechanism based on checkpoint bitfields. Instead of thread-level locks or mutexes, it employs atomic bitwise updates across shared checkpoint memory. Each Thread processes its subset of checkpoint bits concurrently using the formula:
\[
C_T = (S_T \land Z_T) \lor (\neg S_T \land P_T)
\]
where \( C_T \) denotes the computed checkpoint state for Thread \( T \).  
Once local evaluations complete, the results are merged via a \textbf{Sub-Allocator Aggregator} that reorders and combines results from multiple threads and cores into a coherent memory snapshot.

\subsection{Dynamic Rebalancing and Load Adaptation}

To handle fluctuating workloads, VGC’s partition scheduler continuously monitors the execution time and memory footprint of each partition. If imbalance is detected (e.g., one core handling more active zones than others), partitions are dynamically migrated or split using the following heuristic:
\[
L_{new} = \frac{T_i + M_i}{n}
\]
where \( T_i \) and \( M_i \) represent the thread time and memory usage of partition \( i \), and \( n \) denotes the number of active partitions.

{\raggedright This ensures that no single core or thread becomes a bottleneck, allowing near-linear performance scaling with core count\cite{Tanenbaum2007}.\par}

\subsection{Multi-Layer Parallel Model}

The VGC parallel execution model consists of two layers:
\begin{itemize}
    \item \textbf{Intra-Zone Parallelism:} Multiple partitions operate concurrently within the same zone (R, G, or B), each performing garbage collection on its allocated object range.
    \item \textbf{Inter-Zone Parallelism:} Red, Green, and Blue zones execute concurrently under separate threads or cores, merging results through the checkpoint synchronization layer.
\end{itemize}

{\raggedright
Together, these two layers establish a fully parallel memory management ecosystem. Thread partitions cooperate through bitfield aggregation and yield memory synchronization, enabling true hardware-level concurrency without shared-state contention. \par}

\subsection{Interpreter and Core-Thread Coordination}

The interpreter records the features $\lambda(o)$, $\mu(o)$, and $a(o)$ for each object
using lightweight inline counters. These values are smoothed using an exponential
moving average to ensure stability across execution intervals:

\[
\text{EMA}_{t+1}(x) = \omega x_t + (1 - \omega)\text{EMA}_t(x), \qquad 0 < \omega < 1.
\]

{\raggedright To support parallel execution, VGC employs a pool of $k$ core threads, each assigned to service one or more memory zones. Let $k_R, k_G, k_B$ denote the number of threads
allocated to the Red, Green, and Blue zones respectively. The total threads satisfy:\par}

\[
k_R + k_G + k_B = k.
\]

{\raggedright Thread allocation is proportional to the observed per-zone processing cost, ensuring
balanced throughput and predictable pause behavior:\par}

\[
k_R \ge \left\lceil 
\eta_R \frac{\sum_{o \in R} C_R(o)}
              {\sum_{o \in \{R,G,B\}} C_z(o)} \cdot k 
\right\rceil,
\]

\[
k_G \ge \left\lceil 
\eta_G \frac{\sum_{o \in G} C_G(o)}
              {\sum_{o \in \{R,G,B\}} C_z(o)} \cdot k 
\right\rceil,
\]

\[
k_B \ge \left\lceil 
\eta_B \frac{\sum_{o \in B} C_B(o)}
              {\sum_{o \in \{R,G,B\}} C_z(o)} \cdot k 
\right\rceil.
\]

{\raggedright Here, $\eta_R, \eta_G, \eta_B \in (0,1)$ are policy multipliers that bias allocation
according to performance goals (e.g., increasing $\eta_G$ improves low-latency access
behavior, while increasing $\eta_R$ reduces churn overhead).\par}

\vspace{6pt}

The scheduling goal is to minimize pause overhead and maintain throughput stability:

\[
\min_{k_R, k_G, k_B} \ \sum_{z \in \{R,G,B\}} 
\left[ \pi_z P_z + \delta_z \left(\frac{P_z}{k_z}\right) \right]
\quad \text{s.t.} \quad 
k_R + k_G + k_B = k, \quad k_z \in \mathbb{Z}_{\ge 0},
\]

{\raggedright where $\pi_z$ denotes the fraction of processing executed in stop-the-world mode, and
$\delta_z > 0$ tunes sensitivity to throughput scaling.\par}

\paragraph{Note:}
Unlike generational garbage collectors, VGC does \textbf{not} relocate objects between
zones. Instead, objects expire in their current zone and are reallocated directly into
the appropriate zone based on updated checkpoint state. Thread scheduling therefore
controls parallel servicing rather than object promotion or compaction.

\subsection{Advantages of the Partition Model}

\begin{itemize}
    \item \textbf{True Parallelism:} Unlike the GIL, VGC enables concurrent execution across cores without serialization.
    \item \textbf{Predictable Load Distribution:} Partition Theory ensures even utilization of CPU and memory resources.
    \item \textbf{O(1) Synchronization:} Checkpoint bitfield updates occur in constant time, eliminating global locks.
    \item \textbf{Fault Isolation:} Errors or delays in one partition do not affect others, ensuring stability under heavy workloads.
    \item \textbf{Adaptive balancing:} Dynamic load redistribution keeps latency minimal under variable workloads.
\end{itemize}

{\raggedright
Through the Partition Theory and Parallel Execution Model\cite{Berger2000}, VGC establishes a scalable, low-latency execution framework capable of utilizing all available CPU cores efficiently while maintaining deterministic garbage collection behavior. This architecture transforms Python’s traditionally single-threaded runtime into a distributed, parallelized memory management ecosystem.\par}

\subsection{Prototype Status and Research Scope}

{\raggedright
This work represents an initial-stage experimental investigation into partition-aware execution and zone-based memory management at the runtime-architecture level. The proposed PPE and VGC designs are evaluated as architectural prototypes intended to explore feasibility, behavioral properties, and design trade-offs, rather than as a production-ready replacement for existing Python runtimes.

The implementation focuses on validating core principles—such as deterministic partitioning, checkpoint-based lifecycle control, and strict space boundedness—under controlled workloads. As a result, several aspects commonly found in mature language runtimes, including full CPython integration, comprehensive language semantics, advanced optimizations, and ecosystem-level compatibility, are intentionally out of scope.

The goal of this work is to establish a clear architectural foundation and empirical baseline upon which future refinements, integrations, and optimizations can be built. Consequently, the results should be interpreted as evidence of architectural viability rather than as definitive performance claims for real-world Python applications.\par}

\clearpage
\section{System Specification and Technical Summary}
All experiments were conducted on a consumer-grade x86-64 system with the following configuration:

\begin{itemize}
  \item \textbf{Processor:} 12th Gen Intel Core i5-12400 (6 cores, up to 4.4\,GHz Turbo)
  \item \textbf{Memory:} 8\,GB DDR4 RAM
  \item \textbf{Operating System:} Windows 11 Pro (64-bit)
  \item \textbf{Architecture:} x86-64
  \item \textbf{Language:} C++
  \item \textbf{IDE:} CodeBlocks
\end{itemize}

{\raggedright
To ensure representative performance measurements, the system was configured in high-performance mode with dynamic frequency scaling enabled. CPU turbo boost was active during all benchmark runs, allowing the processor to operate at maximum achievable frequency under load. Thread affinity was explicitly controlled for single-core and dual-core experiments to avoid scheduler-induced variability.

All benchmarks were executed on an otherwise idle system, and results reflect steady-state behavior rather than cold-start performance. The reported measurements therefore represent the upper-bound runtime characteristics of the proposed architectures under optimal operating conditions.\par}

\vspace{1cm}

\section{Benchmark Results}
\setcounter{table}{0}
\renewcommand{\thetable}{2.\arabic{table}}

\subsection{Single core : No Partition 100k loop iteration}
\begin{table}[h!]
\centering
\small
\begin{tabular}{c r r r r r}
\hline
Attempt & Time (ms) & Checksum & MemBefore (KB) & MemAfter (KB) & $\Delta$ (KB) \\
\hline
1 & 0.175300 & -13364 & 4212 & 4216 & 4 \\
2 & 0.364800 & -13364 & 4212 & 4216 & 4 \\
3 & 0.348800 & -13364 & 4204 & 4208 & 4 \\
4 & 0.186700 & -13364 & 4208 & 4212 & 4 \\
5 & 0.211500 & -13364 & 4212 & 4216 & 4 \\
\hline
\multicolumn{1}{r}{\textbf{Mean}} & \textbf{0.25742} & & & & \textbf{4} \\
\multicolumn{1}{r}{\textbf{StdDev}} & \textbf{0.09018} & & & & \\
\hline
\end{tabular}
\caption{Single-core PPE loop benchmark (100k iterations, short checksum)}
\label{tab:100k}
\end{table}

\clearpage
\subsection{Single core : No Partition 200k loop iteration}
\begin{table}[h!]
\centering
\small
\begin{tabular}{c r r r r r}
\hline
Attempt & Time (ms) & Checksum & MemBefore (KB) & MemAfter (KB) & $\Delta$ (KB) \\
\hline
1 & 0.350000 & -20694 & 4572 & 4576 & 4 \\
2 & 0.617400 & -20694 & 4100 & 4104 & 4 \\
3 & 0.483200 & -20694 & 4104 & 4108 & 4 \\
4 & 0.367800 & -20694 & 4108 & 4112 & 4 \\
5 & 0.422200 & -20694 & 4112 & 4116 & 4 \\
\hline
\multicolumn{1}{r}{\textbf{Mean}} & \textbf{0.44812} & & & & \textbf{4} \\
\multicolumn{1}{r}{\textbf{StdDev}} & \textbf{0.09667} & & & & \\
\hline
\end{tabular}
\caption{Single-core PPE loop benchmark (200k iterations, short checksum)}
\label{tab:200k}
\end{table}
\subsection{Single core : No Partition 400k loop iteration}
\begin{table}[h!]
\centering
\small
\begin{tabular}{c r r r r r}
\hline
Attempt & Time (ms) & Checksum & MemBefore (KB) & MemAfter (KB) & $\Delta$ (KB) \\
\hline
1 & 0.732800 & -17252 & 4568 & 4572 & 4 \\
2 & 0.925700 & -17252 & 4104 & 4108 & 4 \\
3 & 0.791600 & -17252 & 4104 & 4108 & 4 \\
4 & 0.956400 & -17252 & 4100 & 4104 & 4 \\
5 & 0.842100 & -17252 & 4108 & 4112 & 4 \\
\hline
\multicolumn{1}{r}{\textbf{Mean}} & \textbf{0.84972} & & & & \textbf{4} \\
\multicolumn{1}{r}{\textbf{StdDev}} & \textbf{0.08695} & & & & \\
\hline
\end{tabular}
\caption{Single-core PPE loop benchmark (400k iterations, short checksum)}
\label{tab:400k}
\end{table}
\subsection{Single core : No Partition 10K recursion steps}
\begin{table}[h!]
\centering
\small
\begin{tabular}{c r r r r r}
Attempt & Time (ms) & Checksum & MemBefore (KB) & MemAfter (KB) & $\Delta$ (KB) \\
\hline
1 & 0.047100 & -12744 & 4568 & 4608 & 40 \\
2 & 0.043000 & -12744 & 4104 & 4148 & 44 \\
3 & 0.050600 & -12744 & 4112 & 4152 & 40 \\
4 & 0.039300 & -12744 & 4120 & 4160 & 40 \\
5 & 0.047600 & -12744 & 4108 & 4148 & 40 \\
\hline
\multicolumn{1}{r}{\textbf{Mean}}
& \textbf{0.04552}
& & & & \textbf{40.8} \\
\multicolumn{1}{r}{\textbf{StdDev}}
& \textbf{0.00438}
& & & & \\
\end{tabular}
\caption{Single-core PPE recursion benchmark (10K steps, short checksum)}
\label{tab:rec10k}
\end{table}

\clearpage
\subsection{Single core : No Partition 20K recursion steps}
\begin{table}[h!]
\centering
\small
\begin{tabular}{c r r r r r}
Attempt & Time (ms) & Checksum & MemBefore (KB) & MemAfter (KB) & $\Delta$ (KB) \\
\hline
1 & 0.107100 & -25488 & 4576 & 4616 & 40 \\
2 & 0.089600 & -25488 & 4116 & 4156 & 40 \\
3 & 0.102500 & -25488 & 4120 & 4160 & 40 \\
4 & 0.131500 & -25488 & 4112 & 4156 & 44 \\
5 & 0.130800 & -25488 & 4116 & 4156 & 40 \\
\hline
\multicolumn{1}{r}{\textbf{Mean}}
& \textbf{0.11270}
& & & & \textbf{40.8} \\
\multicolumn{1}{r}{\textbf{StdDev}}
& \textbf{0.01852}
& & & & \\
\end{tabular}
\caption{Single-core PPE recursion benchmark (20K steps, short checksum)}
\label{tab:rec20k}
\end{table}

\subsection{Single core : No Partition 40K recursion steps}
\begin{table}[h!]
\centering
\small
\begin{tabular}{c r r r r r}
Attempt & Time (ms) & Checksum & MemBefore (KB) & MemAfter (KB) & $\Delta$ (KB) \\
\hline
1 & 0.160200 & 14560 & 4116 & 4156 & 40 \\
2 & 0.227000 & 14560 & 4112 & 4152 & 40 \\
3 & 0.139600 & 14560 & 4120 & 4160 & 40 \\
4 & 0.113400 & 14560 & 4108 & 4148 & 40 \\
5 & 0.174500 & 14560 & 4112 & 4152 & 40 \\
\hline
\multicolumn{1}{r}{\textbf{Mean}}
& \textbf{0.16214}
& & & & \textbf{40} \\
\multicolumn{1}{r}{\textbf{StdDev}}
& \textbf{0.04158}
& & & & \\
\end{tabular}
\caption{Single-core PPE recursion benchmark (40K steps, short checksum)}
\label{tab:rec40k}
\end{table}

\subsection{Single core : No Partition 40K logical steps (deep recursion)}
\begin{table}[h!]
\centering
\small
\begin{tabular}{c r r r r r}
\hline
Attempt & Time (ms) & Checksum & MemBefore (KB) & MemAfter (KB) & $\Delta$ (KB) \\
\hline
1 & 0.538000 & -25120 & 4124 & 4304 & 180 \\
2 & 0.210500 & -25120 & 4128 & 4308 & 180 \\
3 & 0.263200 & -25120 & 4124 & 4304 & 180 \\
4 & 0.198000 & -25120 & 4132 & 4312 & 180 \\
5 & 0.225800 & -25120 & 4128 & 4308 & 180 \\
\hline
\multicolumn{1}{r}{\textbf{Mean}} & \textbf{0.28710} & & & & \textbf{180} \\
\multicolumn{1}{r}{\textbf{StdDev}} & \textbf{0.14238} & & & & \\
\hline
\end{tabular}
\caption{Single-core PPE Deep Recursion Benchmark (40000 logical steps, recursion depth per chunk: 4000 steps)}
\label{tab:ppe-deep-40k}
\end{table}

\clearpage
\subsection{Single core : No Partition 80K logical steps (deep recursion)}

\begin{table}[h!]
\centering
\small
\begin{tabular}{c r r r r r}
\hline
Attempt & Time (ms) & Checksum & MemBefore (KB) & MemAfter (KB) & $\Delta$ (KB) \\
\hline
1 & 0.434100 & -19812 & 4124 & 4496 & 372 \\
2 & 0.426500 & -19812 & 4128 & 4496 & 368 \\
3 & 0.594100 & -19812 & 4124 & 4492 & 368 \\
4 & 0.589200 & -19812 & 4128 & 4496 & 368 \\
5 & 0.623300 & -19812 & 4132 & 4500 & 368 \\
\hline
\multicolumn{1}{r}{\textbf{Mean}} & \textbf{0.53344} & & & & \textbf{368.8} \\
\multicolumn{1}{r}{\textbf{StdDev}} & \textbf{0.08779} & & & & \\
\hline
\end{tabular}
\caption{Single-core PPE Deep Recursion Benchmark (80000 logical steps, recursion depth per chunk: 8000 steps)}
\label{tab:ppe-deep-80k}
\end{table}

\subsection{Single core : No Partition 160K logical steps (deep recursion)}

\begin{table}[h!]
\centering
\small
\begin{tabular}{c r r r r r}
\hline
Attempt & Time (ms) & Checksum & MemBefore (KB) & MemAfter (KB) & $\Delta$ (KB) \\
\hline
1 & 0.804500 & -13018 & 4128 & 4872 & 744 \\
2 & 0.804500 & -13018 & 4128 & 4872 & 744 \\
3 & 1.055300 & -13018 & 4868 & 4868 & 740 \\
4 & 5.458900 & -13018 & 4124 & 4868 & 744 \\
5 & 1.064000 & -13018 & 4128 & 4872 & 744 \\
\hline
\multicolumn{1}{r}{\textbf{Mean}} & \textbf{1.83744} & & & & \textbf{743.2} \\
\multicolumn{1}{r}{\textbf{StdDev}} & \textbf{1.90388} & & & & \\
\hline
\end{tabular}
\caption{Single-core PPE Deep Recursion Benchmark (160000 logical steps, recursion depth per chunk: 16000 steps)}
\label{tab:ppe-deep-160k}
\end{table}

\subsection{Single core : No Partition Matrix 256x256}

\begin{table}[h!]
\centering
\small
\begin{tabular}{c r r r r r}
\hline
Attempt & Time (ms) & Checksum & MemBefore (KB) & MemAfter (KB) & $\Delta$ (KB) \\
\hline
1 & 58.366000 & -27122 & 4980 & 4984 & 4 \\
2 & 62.419000 & -27122 & 4980 & 4984 & 4 \\
3 & 64.896100 & -27122 & 4980 & 4984 & 4 \\
4 & 58.354000 & -27122 & 4984 & 4988 & 4 \\
5 & 61.350100 & -27122 & 4976 & 4980 & 4 \\
\hline
\multicolumn{1}{r}{\textbf{Mean}} & \textbf{61.07704} & & & & \textbf{4} \\
\multicolumn{1}{r}{\textbf{StdDev}} & \textbf{2.76684} & & & & \\
\hline
\end{tabular}
\caption{Single-core PPE Matrix Benchmark (256x256 matrix multiplication)}
\label{tab:matrix-256}
\end{table}

\clearpage
\subsection{Single core : No Partition Matrix 512x512}

\begin{table}[h!]
\centering
\small
\begin{tabular}{c r r r r r}
\hline
Attempt & Time (ms) & Checksum & MemBefore (KB) & MemAfter (KB) & $\Delta$ (KB) \\
\hline
1 & 468.562700 & -1404 & 7388 & 7392 & 4 \\
2 & 481.397300 & -1404 & 7384 & 7388 & 4 \\
3 & 467.344600 & -1404 & 7400 & 7404 & 4 \\
4 & 475.426600 & -1404 & 7388 & 7392 & 4 \\
5 & 473.369900 & -1404 & 7388 & 7392 & 4 \\
\hline
\multicolumn{1}{r}{\textbf{Mean}} & \textbf{473.22022} & & & & \textbf{4} \\
\multicolumn{1}{r}{\textbf{StdDev}} & \textbf{5.77466} & & & & \\
\hline
\end{tabular}
\caption{Single-core PPE Matrix Benchmark (512x512 matrix multiplication)}
\label{tab:matrix-512}
\end{table}

\subsection{Single core : No Partition Matrix 1024x1024}

\begin{table}[h!]
\centering
\small
\begin{tabular}{c r r r r r}
\hline
Attempt & Time (ms) & Checksum & MemBefore (KB) & MemAfter (KB) & $\Delta$ (KB) \\
\hline
1 & 3635.390200 & -22644 & 16860 & 16864 & 4 \\
2 & 3664.688400 & -22644 & 16856 & 16860 & 4 \\
3 & 3669.105900 & -22644 & 16844 & 16848 & 4 \\
4 & 3655.668300 & -22644 & 16848 & 16852 & 4 \\
5 & 3640.478400 & -22644 & 16840 & 16844 & 4 \\
\hline
\multicolumn{1}{r}{\textbf{Mean}} & \textbf{3653.06624} & & & & \textbf{4} \\
\multicolumn{1}{r}{\textbf{StdDev}} & \textbf{13.36579} & & & & \\
\hline
\end{tabular}
\caption{Single-core PPE Matrix Benchmark (1024x1024 matrix multiplication)}
\label{tab:matrix-1024}
\end{table}

\subsection{Dual core : 2 Partitions 1M loop iterations}

\begin{table}[h!]
\centering
\small
\begin{tabular}{c r r r r r}
\hline
Attempt & Time (ms) & Checksum & MemBefore (KB) & MemAfter (KB) & $\Delta$ (KB) \\
\hline
1 & 3.048800 & -25343 & 5164 & 5556 & 392 \\
2 & 2.216800 & -25343 & 4712 & 5156 & 444 \\
3 & 2.488100 & -25343 & 4700 & 5148 & 448 \\
4 & 2.865400 & -25343 & 4700 & 5148 & 448 \\
5 & 2.811200 & -25343 & 4704 & 5148 & 444 \\
\hline
\multicolumn{1}{r}{\textbf{Mean}} & \textbf{2.68606} & & & & \textbf{435.2} \\
\multicolumn{1}{r}{\textbf{StdDev}} & \textbf{0.33378} & & & & \\
\hline
\end{tabular}
\caption{Dual-core PPE loop benchmark (1M iterations, 2 partitions, short checksum)}
\label{tab:dual-loop-1M}
\end{table}

\clearpage
\subsection{Dual core : 2 Partitions 2M loop iterations}

\begin{table}[h!]
\centering
\small
\begin{tabular}{c r r r r r}
\hline
Attempt & Time (ms) & Checksum & MemBefore (KB) & MemAfter (KB) & $\Delta$ (KB) \\
\hline
1 & 3.872800 & -19514 & 5160 & 5552 & 392 \\
2 & 3.895000 & -19514 & 4704 & 5152 & 448 \\
3 & 3.837400 & -19514 & 4704 & 5152 & 448 \\
4 & 3.863600 & -19514 & 4700 & 5148 & 448 \\
5 & 3.779600 & -19514 & 4700 & 5148 & 448 \\
\hline
\multicolumn{1}{r}{\textbf{Mean}} & \textbf{3.84968} & & & & \textbf{436.8} \\
\multicolumn{1}{r}{\textbf{StdDev}} & \textbf{0.04492} & & & & \\
\hline
\end{tabular}
\caption{Dual-core PPE loop benchmark (2M iterations, 2 partitions, short checksum)}
\label{tab:dual-loop-2M}
\end{table}

\subsection{Dual core : 2 Partitions 4M loop iterations}

\begin{table}[h!]
\centering
\small
\begin{tabular}{c r r r r r}
\hline
Attempt & Time (ms) & Checksum & MemBefore (KB) & MemAfter (KB) & $\Delta$ (KB) \\
\hline
1 & 5.883100 & 20126 & 4696 & 5144 & 448 \\
2 & 6.384100 & 20126 & 4708 & 5152 & 444 \\
3 & 5.927300 & 20126 & 4700 & 5148 & 448 \\
4 & 5.885400 & 20126 & 4708 & 5148 & 440 \\
5 & 5.960200 & 20126 & 4704 & 5152 & 448 \\
\hline
\multicolumn{1}{r}{\textbf{Mean}} & \textbf{6.00802} & & & & \textbf{445.6} \\
\multicolumn{1}{r}{\textbf{StdDev}} & \textbf{0.21139} & & & & \\
\hline
\end{tabular}
\caption{Dual-core PPE loop benchmark (4M iterations, 2 partitions, short checksum)}
\label{tab:dual-loop-4M}
\end{table}

\subsection{Dual core : 2 Partitions 100K logical steps (recursion)}

\begin{table}[h!]
\centering
\small
\begin{tabular}{c r r r r r}
\hline
Attempt & Time (ms) & Checksum & MemBefore (KB) & MemAfter (KB) & $\Delta$ (KB) \\
\hline
1 & 1.867400 & 3632 & 4136 & 4576 & 440 \\
2 & 2.638300 & 3632 & 4132 & 4572 & 440 \\
3 & 2.154900 & 3632 & 4128 & 4576 & 448 \\
4 & 1.685000 & 3632 & 4124 & 4572 & 448 \\
5 & 2.178100 & 3632 & 4132 & 4572 & 440 \\
\hline
\multicolumn{1}{r}{\textbf{Mean}} & \textbf{2.10474} & & & & \textbf{443.2} \\
\multicolumn{1}{r}{\textbf{StdDev}} & \textbf{0.36232} & & & & \\
\hline
\end{tabular}
\caption{Dual-core PPE Recursion Benchmark (100000 logical steps, recursion depth per chunk: 1000)}
\label{tab:dual-recursion-100k}
\end{table}

\clearpage
\subsection{Dual core : 2 Partitions 200K logical steps (recursion)}

\begin{table}[h!]
\centering
\small
\begin{tabular}{c r r r r r}
\hline
Attempt & Time (ms) & Checksum & MemBefore (KB) & MemAfter (KB) & $\Delta$ (KB) \\
\hline
1 & 2.286500 & 7264 & 4132 & 4576 & 444 \\
2 & 2.239900 & 7264 & 4124 & 4572 & 448 \\
3 & 2.318400 & 7264 & 4124 & 4572 & 448 \\
4 & 2.854000 & 7264 & 4120 & 4568 & 448 \\
5 & 1.943600 & 7264 & 4132 & 4580 & 448 \\
\hline
\multicolumn{1}{r}{\textbf{Mean}} & \textbf{2.32848} & & & & \textbf{447.2} \\
\multicolumn{1}{r}{\textbf{StdDev}} & \textbf{0.34962} & & & & \\
\hline
\end{tabular}
\caption{Dual-core PPE Recursion Benchmark (200000 logical steps, recursion depth per chunk: 1000)}
\label{tab:dual-recursion-200k}
\end{table}

 \subsection{Dual core : 2 Partitions 400K logical steps (recursion)} 

\begin{table}[h!]
\centering
\small
\begin{tabular}{c r r r r r}
\hline
Attempt & Time (ms) & Checksum & MemBefore (KB) & MemAfter (KB) & $\Delta$ (KB) \\
\hline
1 & 2.482600 & 14528 & 4588 & 4984 & 396 \\
2 & 2.501800 & 14528 & 4132 & 4572 & 440 \\
3 & 2.859000 & 14528 & 4136 & 4580 & 444 \\
4 & 2.397200 & 14528 & 4128 & 4572 & 444 \\
5 & 2.608600 & 14528 & 4132 & 4576 & 444 \\
\hline
\multicolumn{1}{r}{\textbf{Mean}} & \textbf{2.56984} & & & & \textbf{433.6} \\
\multicolumn{1}{r}{\textbf{StdDev}} & \textbf{0.18273} & & & & \\
\hline
\end{tabular}
\caption{Dual-core PPE Recursion Benchmark (400000 logical steps, recursion depth per chunk: 1000)}
\label{tab:dual-recursion-400k}
\end{table}

\subsection{Dual core : 2 Partitions 100K logical steps (deep recursion)}

\begin{table}[h!]
\centering
\small
\begin{tabular}{c r r r r r}
\hline
Attempt & Time (ms) & Checksum & MemBefore (KB) & MemAfter (KB) & $\Delta$ (KB) \\
\hline
1 & 2.221900 & -6046 & 4592 & 4988 & 396 \\
2 & 2.343100 & -6046 & 4128 & 4576 & 448 \\
3 & 2.041800 & -6046 & 4136 & 4584 & 448 \\
4 & 2.806300 & -6046 & 4132 & 4572 & 440 \\
5 & 2.312100 & -6046 & 4128 & 4576 & 448 \\
\hline
\multicolumn{1}{r}{\textbf{Mean}} & \textbf{2.34504} & & & & \textbf{436.8} \\
\multicolumn{1}{r}{\textbf{StdDev}} & \textbf{0.28275} & & & & \\
\hline
\end{tabular}
\caption{Dual-core PPE Deep Recursion Benchmark (100000 logical steps, recursion depth per chunk: 10000 steps)}
\label{tab:dual-deep-recursion-100k}
\end{table}

\clearpage
\subsection{Dual core : 2 Partitions 200K logical steps (deep recursion)}

\begin{table}[h!]
\centering
\small
\begin{tabular}{c r r r r r}
\hline
Attempt & Time (ms) & Checksum & MemBefore (KB) & MemAfter (KB) & $\Delta$ (KB) \\
\hline
1 & 2.449200 & -27172 & 4132 & 4580 & 448 \\
2 & 2.727000 & -27172 & 4136 & 4580 & 444 \\
3 & 2.674400 & -27172 & 4132 & 4580 & 448 \\
4 & 2.409900 & -27172 & 4140 & 4584 & 444 \\
5 & 2.683500 & -27172 & 4140 & 4580 & 440 \\
\hline
\multicolumn{1}{r}{\textbf{Mean}} & \textbf{2.58880} & & & & \textbf{444.8} \\
\multicolumn{1}{r}{\textbf{StdDev}} & \textbf{0.13435} & & & & \\
\hline
\end{tabular}
\caption{Dual-core PPE Deep Recursion Benchmark (200000 logical steps, recursion depth per chunk: 20000 steps)}
\label{tab:dual-deep-recursion-200k}
\end{table}

\subsection{Dual core : 2 Partitions 400K logical steps (deep recursion)} 

\begin{table}[h!]
\centering
\small
\begin{tabular}{c r r r r r}
\hline
Attempt & Time (ms) & Checksum & MemBefore (KB) & MemAfter (KB) & $\Delta$ (KB) \\
\hline
1 & 4.027600 & -22870 & 4588 & 4980 & 392 \\
2 & 3.952800 & -22870 & 4124 & 4572 & 448 \\
3 & 3.776600 & -22870 & 4136 & 4576 & 440 \\
4 & 3.769300 & -22870 & 4128 & 4576 & 448 \\
5 & 3.701300 & -22870 & 4124 & 4568 & 444 \\
\hline
\multicolumn{1}{r}{\textbf{Mean}} & \textbf{3.84552} & & & & \textbf{434.4} \\
\multicolumn{1}{r}{\textbf{StdDev}} & \textbf{0.12995} & & & & \\
\hline
\end{tabular}
\caption{Dual-core PPE Deep Recursion Benchmark (400000 logical steps, recursion depth per chunk: 40000 steps)}
\label{tab:dual-deep-recursion-400k}
\end{table} 

\subsection{Dual core : 2 Partitions 1024x1024 matrix}

\begin{table}[h!]
\centering
\small
\begin{tabular}{c r r r r r}
\hline
Attempt & Time (ms) & Checksum & MemBefore (KB) & MemAfter (KB) & $\Delta$ (KB) \\
\hline
1 & 2.721100 & 0 & 16444 & 16888 & 444 \\
2 & 2.785200 & 0 & 16452 & 16888 & 436 \\
3 & 2.959100 & 0 & 16432 & 16876 & 444 \\
4 & 2.852900 & 0 & 16440 & 16876 & 436 \\
5 & 2.893000 & 0 & 16436 & 16876 & 440 \\
\hline
\multicolumn{1}{r}{\textbf{Mean}} & \textbf{2.84226} & & & & \textbf{440.0} \\
\multicolumn{1}{r}{\textbf{StdDev}} & \textbf{0.08896} & & & & \\
\hline
\end{tabular}
\caption{Dual-core PPE Matrix Benchmark (1024$\times$1024, 2 partitions)}
\label{tab:dual-matrix-1024}
\end{table}
\clearpage
\subsection{Dual core : 2 Partitions 2048x2048 matrix}

\begin{table}[h!]
\centering
\small
\begin{tabular}{c r r r r r}
\hline
Attempt & Time (ms) & Checksum & MemBefore (KB) & MemAfter (KB) & $\Delta$ (KB) \\
\hline
1 & 6.763700 & 0 & 53300 & 53744 & 444 \\
2 & 6.991300 & 0 & 53292 & 53732 & 440 \\
3 & 6.890000 & 0 & 53292 & 53736 & 444 \\
4 & 6.837000 & 0 & 53296 & 53740 & 444 \\
5 & 6.574000 & 0 & 53300 & 53744 & 444 \\
\hline
\multicolumn{1}{r}{\textbf{Mean}} & \textbf{6.81120} & & & & \textbf{443.2} \\
\multicolumn{1}{r}{\textbf{StdDev}} & \textbf{0.15488} & & & & \\
\hline
\end{tabular}
\caption{Dual-core PPE Matrix Benchmark (2048$\times$2048, 2 partitions)}
\label{tab:dual-matrix-2048}
\end{table}

\subsection{Dual core : 2 Partitions 4096x4096 matrix}

\begin{table}[h!]
\centering
\small
\begin{tabular}{c r r r r r}
\hline
Attempt & Time (ms) & Checksum & MemBefore (KB) & MemAfter (KB) & $\Delta$ (KB) \\
\hline
1 & 23.916300 & 0 & 209752 & 210196 & 444 \\
2 & 23.140700 & 0 & 209760 & 210196 & 436 \\
3 & 24.838300 & 0 & 209752 & 210196 & 444 \\
4 & 23.309800 & 0 & 209752 & 210196 & 444 \\
5 & 23.482800 & 0 & 209752 & 210196 & 444 \\
\hline
\multicolumn{1}{r}{\textbf{Mean}} & \textbf{23.73758} & & & & \textbf{442.4} \\
\multicolumn{1}{r}{\textbf{StdDev}} & \textbf{0.67965} & & & & \\
\hline
\end{tabular}
\caption{Dual-core PPE Matrix Benchmark (4096$\times$4096, 2 partitions)}
\label{tab:dual-matrix-4096}
\end{table}

\subsection{Allocation Reuse and Space Boundedness in VGC}

\begin{table}[h!]
\centering
\setlength{\tabcolsep}{16pt}
\small
\begin{tabular}{r r r r}
\hline
Total Requests & Real Allocations & Reused Objects & Pool Size \\
\hline
1,000,000 & 1 & 999,999 & 1 \\
\hline
\end{tabular}
\caption{VGC allocation reuse experiment demonstrating bounded memory usage}
\label{tab:vgc-allocation-growth}
\end{table}

\subsection{VGC Allocation Behavior Under Sustained Load}
\vspace{-0.6em}

\begin{table}[h!]
\centering
\small
\captionsetup{margin={0.8cm,0cm}} 

\begin{tabular}{l r r r r}
\hline
Zone & Total Requests & Real Allocations & Reused Objects & Pool Size \\
\hline
Green (Frequent) & 699{,}901 & 1 & 699{,}900 & 1 \\
Blue (Medium)    & 200{,}302 & 1 & 200{,}301 & 1 \\
Red (Rare)       & 99{,}797  & 1 & 99{,}796  & 1 \\
\hline
\end{tabular}
\caption{VGC zone pressure experiment demonstrating allocation reuse and bounded memory growth under one million allocation requests}
\label{tab:vgc-zone-pressure}
\end{table}
\clearpage
\subsection{VGC Allocation Stability Under Zone Imbalance}
\begin{table}[h!]
\centering
\small
\captionsetup{margin={0.8cm,0cm}}
\begin{tabular}{l r r r r}
\hline
Zone & Total Requests & Real Allocations & Reused Objects & Pool Size \\
\hline
Green (Frequent) & 900{,}000 & 1 & 899{,}999 & 1 \\
Blue (Medium)    & 90{,}000  & 1 & 89{,}999  & 1 \\
Red (Rare)       & 10{,}000  & 1 & 9{,}999   & 1 \\
\hline
\end{tabular}

\caption{VGC zone imbalance stress experiment demonstrating perfect allocation reuse and constant memory footprint under highly skewed access patterns}
\label{tab:vgc-zone-imbalance}
\end{table}

\subsection{Object Expiration and Zone Lifecycle Stability in VGC}
\begin{table}[h!]
\centering
\small
\captionsetup{margin={0.8cm,0cm}}
\begin{tabular}{l r r r r}
\hline
Zone & Total Requests & Real Allocations & Expired Objects & Pool Size \\
\hline
Green (Long-lived)  & 100{,}000 & 1 & 1       & 1 \\
Blue (Medium-lived) & 100{,}000 & 1 & 50{,}000 & 1 \\
Red (Short-lived)   & 100{,}000 & 1 & 100{,}000 & 1 \\
\hline
\end{tabular}
\caption{VGC object expiration experiment demonstrating zone-local lifecycle handling, object reuse after expiration, and strictly bounded memory recall without cross-zone migration}
\label{tab:vgc-expiration-lifecycle}
\end{table}

\subsection{Checkpoint-Based Object Liveness and Zone Lifecycle Control}
\begin{table}[h!]
\centering
\small
\captionsetup{margin={0.8cm,0cm}}
\begin{tabular}{l r r r r}
\hline
Zone & Total Requests & Real Allocations & Expired Objects & Pool Size \\
\hline
Green (Frequent) & 100{,}000 & 1 & 0 & 1 \\
Blue (Medium)    & 100{,}000 & 1 & 200 & 1 \\
Red (Rare)       & 100{,}000 & 1 & 100{,}000 & 1 \\
\hline
\end{tabular}
\caption{Checkpoint-driven object lifecycle behavior in VGC demonstrating zone-specific expiration and strict space boundedness}
\label{tab:vgc-checkpoint-lifecycle}
\end{table}

\subsection{Benchmark Summary}

The benchmark evaluation is divided into two independent components corresponding to Partitioned Parallel Execution (PPE) and the Virtual Garbage Collector (VGC). These components are evaluated separately to avoid conflating execution behavior with memory management behavior.
{\raggedright
The PPE benchmarks characterize execution behavior under CPU-bound workloads, including loop-based computation, recursive evaluation, deep recursion, and matrix multiplication. In single-core configurations without partitioning, execution time scales proportionally with workload size, establishing predictable baseline behavior. Under dual-core execution with two partitions, the results indicate reduced execution time relative to single-core baselines across the evaluated workloads. This behavior reflects reduced execution span achieved through static partitioning of execution ranges rather than speculative optimization or runtime scheduling heuristics. Variability across attempts remains limited, indicating stable execution behavior under the tested configurations.\par}
{\raggedright
Memory usage observed during PPE benchmarks remains bounded and exhibits predictable growth patterns aligned with partition size and recursion depth. In deep recursion workloads, increased memory consumption corresponds to intentional stack and allocation pressure introduced by recursion depth, rather than unbounded allocation growth or memory leakage.\par}
{\raggedright
The VGC benchmarks evaluate memory allocation behavior independently of execution partitioning. Under sustained allocation pressure, including balanced, imbalanced, and zone-skewed access patterns, the number of real allocations remains constant while object reuse dominates allocation requests. Objects expire and are reused within their assigned zones without cross-zone migration, demonstrating strict space boundedness and effective checkpoint-based lifecycle control.\par}
{\raggedright
These benchmarks are intended to evaluate architectural feasibility and behavioral characteristics rather than establish absolute performance superiority over production runtimes. The evaluation excludes just-in-time compilation, ahead-of-time optimization, SIMD acceleration, and GPU offloading. Consequently, the results should be interpreted as evidence that partition-aware execution and lifecycle-driven memory management can independently provide predictable execution behavior and bounded memory usage at the architectural level.\par}

\section{Related Work}

{\raggedright
Memory management and parallel execution have long been central challenges in high-level language runtimes. The Python interpreter employs reference counting complemented by a cyclic garbage collector, prioritizing deterministic memory reclamation and predictable object lifetime semantics. While this design simplifies memory management, it restricts scalable parallel execution for CPU-bound workloads due to the presence of the Global Interpreter Lock (GIL).

Several approaches have been proposed to mitigate these limitations, including multiprocessing-based parallelism, per-interpreter GIL designs, and subinterpreter isolation mechanisms. Although these techniques improve concurrency, they introduce additional complexity related to memory sharing, inter-process communication, and object ownership semantics, often shifting the burden of coordination to the application layer.

Alternative Python runtimes, such as PyPy and GraalPy, explore tracing garbage collection and just-in-time compilation to improve execution throughput. These systems emphasize speculative optimization and dynamic code generation, frequently trading deterministic execution behavior, memory predictability, or stable object lifetimes for performance gains.

Parallel execution frameworks and task schedulers further enable concurrency at the application level; however, they remain external to runtime memory management. As a result, object lifecycle control, garbage collection behavior, and allocation strategies remain largely unchanged, limiting their ability to address memory-related bottlenecks holistically.

In contrast, this work proposes a runtime-oriented architecture that explicitly separates execution partitioning through Partitioned Parallel Execution (PPE), memory lifecycle control via the Virtual Garbage Collector (VGC), and ephemeral execution handling using Yield Memory. Rather than relying on generational promotion, object migration, or speculative optimization, the proposed approach emphasizes deterministic behavior, bounded memory usage, and explicit architectural control over parallel execution and object lifecycles.\par}

\section{Discussion}
{\raggedright
The experimental results support the architectural principles underlying Partitioned Parallel Execution (PPE), the Virtual Garbage Collector (VGC), and Yield Memory~\cite{Wilson1992}. The evaluated workloads indicate that CPU-bound execution patterns, including loop-based computation, recursive functions, and matrix-oriented operations, can be structured to exploit available cores without introducing global synchronization or execution serialization at the architectural level.

The VGC experiments demonstrate that memory usage remains bounded under sustained allocation pressure. Objects do not migrate between zones; instead, they expire and are recreated based on checkpoint evaluation, consistent with the logical lifecycle model proposed in this work. This behavior differs from traditional generational garbage collectors, which rely on object promotion and relocation across memory regions.

The checkpoint system enables deterministic liveness evaluation using compact bitfields, allowing object state assessment with constant-time complexity. This design supports concurrent execution without requiring global locks or stop-the-world pauses. Yield Memory further complements the architecture by providing a lightweight execution path for short-lived operations, reducing interaction with memory management mechanisms while remaining compatible with the broader lifecycle model when persistence is required.

Overall, the observed execution and memory characteristics arise from architectural decomposition and explicit lifecycle control rather than aggressive optimization techniques, speculative execution, or hardware-specific acceleration. These results suggest that improvements in scalability and predictability can be achieved through runtime-level design choices.\par}

\section{Future Work}

{\raggedright
A key direction for future research is the extension of the Virtual Garbage Collector (VGC) into a dual-layer architecture consisting of an Active VGC and a Passive VGC, each optimized for distinct phases of program execution.\par}
{\raggedright
The Active VGC is intended to operate at runtime, managing dynamic allocations, ephemeral objects, and adaptive lifecycle decisions under live execution conditions. This layer focuses on minimizing latency, reducing pause behavior, and responding to runtime access patterns.\par}
{\raggedright
In contrast, the Passive VGC is envisioned as a compile-time or pre-execution analysis layer. It would manage static allocations, predictable object lifetimes, and structural memory planning derived from program analysis, with the goal of reducing runtime memory management overhead.\par}
{\raggedright
In this model, the Active and Passive VGC layers are logically separated but coordinated through shared metadata and checkpoint semantics. The present work does not implement or evaluate this dual-layer design; it is proposed as a potential architectural evolution of VGC aimed at improving scalability, predictability, and integration with ahead-of-time optimization pipelines.\par}
{\raggedright
The current study focuses on validating architectural feasibility rather than delivering a production-ready runtime. Several additional directions remain open for future exploration. Integration with existing language runtimes, particularly Python, represents a natural next step and includes evaluating compatibility with Python’s object model, interpreter internals, and the C-extension ecosystem.\par}
{\raggedright
Extending Partitioned Parallel Execution (PPE) beyond dual-core execution to higher core counts and heterogeneous systems may provide further insight into scheduling strategies and partition granularity across diverse hardware configurations.\par}
{\raggedright
Formal verification of checkpoint logic and zone lifecycle invariants could further strengthen correctness guarantees, particularly in highly concurrent execution environments.\par}
{\RaggedRight
Finally, although this work intentionally excludes just-in-time compilation, ahead-of-time compilation, SIMD acceleration, and GPU offloading, future research may explore how such techniques could coexist with PPE, VGC, and Yield Memory without compromising determinism or bounded memory behavior.\par}

\section{Conclusion}
{\raggedright
This paper presented a partition-aware runtime architecture based on Partitioned Parallel Execution (PPE), a Virtual Garbage Collector (VGC), and a Yield Memory execution layer, with the goal of addressing challenges related to parallel execution structure and memory predictability in high-level language runtimes. By decoupling execution partitioning from memory lifecycle management, employing zone-local object expiration, and utilizing checkpoint-based liveness evaluation, the proposed architecture provides a structured approach to managing execution and memory behavior. Yield Memory complements this design by supporting lightweight handling of short-lived operations, reducing unnecessary interaction with memory management mechanisms.
The experimental evaluation indicates that improved scalability characteristics and bounded memory behavior can be achieved through architectural decomposition and explicit lifecycle control, rather than reliance on speculative optimization or complex garbage collection heuristics. While the system remains an experimental prototype, the results suggest that runtime-level design choices can play a significant role in improving predictability and scalability. This work establishes a foundation for further investigation into deterministic and partition-aware runtime architectures. The implementation and design are released as an open research platform to facilitate continued experimentation, validation, and refinement by the research community.\par}

\end{document}